# Practical Algorithmic Techniques for Several String Processing Problems

Mugurel Ionuţ Andreica, Nicolae Ţăpuş
Computer Science and Engineering Department
Politehnica University of Bucharest
Bucharest, Romania
e-mail: {mugurel.andreica, nicolae.tapus}@cs.pub.ro

*Abstract*—The domains of data mining and knowledge discovery make use of large amounts of textual data, which need to be handled efficiently. Specific problems, like finding the maximum weight ordered common subset of a set of ordered sets or searching for specific patterns within texts, occur frequently in this context. In this paper we present several novel and practical algorithmic techniques for processing textual data (strings) in order to efficiently solve multiple problems. Our techniques make use of efficient string algorithms and data structures, like KMP, suffix arrays, tries and deterministic finite automata.

*Keywords-string processing; prefix query; trie; suffix array; KMP; deterministic finite automaton*

## I. INTRODUCTION

Textual data exists and is constantly being produced in large amounts – web pages, digital libraries or official state documents all need to be processed efficiently in order to be able to extract automatically the information contained within them. Web search engines classify and store the text data into efficient data structures, data mining algorithms make use of similarities between multiple pieces of data, and biologists analyze large DNA sequences in order to extract information regarding genes and identify patterns. In this paper we present novel and practical algorithmic solutions for several string processing problems. Our techniques make use of efficient string algorithms and data structures, like KMP, suffix arrays, tries, deterministic finite automata, and others. In Section II we discuss two types of string prefix queries. In Section III we address the problem of optimally concatenating a set of strings, in order to optimize an objective metric. In Section IV we study constrained optimal length common subsequences and we present novel solutions for the shortest common contiguous non-subsequence of a set of strings. In Section V we construct and count strings having specific properties. Finally, in Section VI we discuss related work and in Section VII we conclude.

## II. STRING PREFIX QUERIES

We consider a string $S=c_1c_2...c_n$ (of length $n$). We want to preprocess the string in order to be able to answer the following types of queries: *1)* $PQ(i,j)$=is the prefix $c_1c_2...c_i$ of $S$ equal to $c_{j-i+1}c_{j-i+2}...c_j$ ? ($j \geq i$) ; *2)* $LPQ(j,k)$=which is the largest value of $i$ such that $i \leq k$ ($0 \leq i \leq k \leq j$) and $PQ(i,j)$=true ? The solution for both types of queries consists of first running the preprocessing stage of the *Knuth-Morris-Pratt* (*KMP*) algorithm [4]. As a result of this stage, the table $p(i)$ is computed, where $p(i)$ is the largest value ($p(i)<i$) such that $PQ(p(i),i)$=*true*. We will construct a tree having $n+1$ vertices (one for each prefix length of $S$: from $0$ to $n$). The parent of the vertex $i$ will be $p(i)$. Vertex $0$ will be the root of the tree. The two types of queries can be translated in terms of this tree (called the *failure tree* [1]). $PQ(i,j)$ returns *true* only if $i$ is an ancestor of the vertex $j$ (including $j$). The answer to $LPQ(j,k)$ is the lowest ancestor $i$ of vertex $j$ such that $i \leq k$. We will describe next efficient solutions for answering these ancestor queries. We perform a DFS traversal of the tree starting from the root (vertex $0$). We assign to each vertex $i$ its DFS number, $DFSnum(i)$. During the traversal, we maintain a counter $q$ (initially, $q=0$). When we first enter a vertex $i$, we increment $q$ by $1$ and then we set $DFSnum(i)=q$. When we exit a vertex $i$ (after visiting its entire subtree) we set $DFSmax(i)=q$. A vertex $i$ is an ancestor of a vertex $j$ if $DFSnum(i) \leq DFSnum(j)$ and $DFSmax(i) \geq DFSmax(j)$.

In order to answer the second type of queries, we compute for each vertex $i$ the values $Anc(i,j)$=the ancestor of vertex $i$ which is located $2^j$ levels higher than $i$. $Anc(i,0)=p(i)$ ($Anc(root,0)=root$) and $Anc(i,j \geq 1)=Anc(Anc(i,j-1),j-1)$. Let's assume that every vertex $v$ has a weight $w(v)$ and we have $w(v) \geq w(p(v))$ (in our case, we have $w(v)=v$). We can find the vertex $v$ with the largest value $w(v)$ which is an ancestor of a vertex $i$ and such that $w(v) \leq k$ (we assume that such a vertex always exists; if not, we will return the root of the tree) in $O(log(n))$ time as follows. We set $j=ceil(log_2(n))$ (we denote by $ceil(x)$ the integer which is equal to $x$ rounded up) and $v=i$. While $j>0$ do: *(1)* if $(w(Anc(v,j))>k)$ then $v=Anc(v,j)$; *(2)* $j=j-1$. In the end, if $w(v)>k$ (and $v$ is not the tree root) we set $v=p(v)$. This approach has the disadvantage that it requires $O(n \cdot log(n))$ memory. We can trade memory for running time, as follows. We will compute a value $Anc(i)$=an ancestor of vertex $i$ which is located $c$ levels higher than $i$ (or the closest ancestor which is located on a level which is a multiple of $c$). While traversing the tree, we maintain the DFS stack $Stk$ of vertices. When first entering a vertex $i$, this vertex is pushed at the top of the stack and when exiting a vertex $i$, this vertex is popped from the stack. Let's assume that the topmost level of the stack is $ltop$ (the one containing vertex $i$). If $ltop \leq c$ we have $Anc(i)$=the tree's root; otherwise, we may set $Anc(i)=Stk(ltop-c)$ or $Anc(i)=Stk(ltop-1-((ltop-1) \bmod c))$. Then, in order to find the ancestor $v$ of vertex $i$ with the largest weight $w(v) \leq k$ we proceed as follows. We initialize $v=i$. While ($v$ is not the tree root) and $w(Anc(v))>k$ we set $v=Anc(v)$. Then, while ($v$ is not the tree root) and $(w(p(v))>k)$ we set $v=p(v)$. In the end, if $w(v)>k$ (and $v$ is not the tree root) we set $v=p(v)$.

## III. OPTIMAL STRING CONCATENATIONS

### A. Optimal Concatenation of Strings from Two Sets

We consider two sets $A$ and $B$, each containing $N$ strings. We want to compute the shortest string $S$ which can be obtained as a concatenation of both some strings from $A$ and of some strings from $B$. When considering the concatenation of the strings from each set, a string may occur zero, one, or multiple times. We will denote by $len(x,y)$ the length of the string $x$ from the set $A$ (if $y=1$) or $B$ (if $y=2$). We will conceptually construct two strings $S_1$ and $S_2$, where $S_1$ is obtained by concatenating some strings from $A$ and $S_2$ is obtained by concatenating some strings from $B$. Let's assume that, initially, $S_1$ contains a string from $A$ and $S_2$ is empty. The algorithm for constructing the strings $S_1$ and $S_2$ will add a string from $A$ ($B$) to $S_1$ ($S_2$) if $S_1$ ($S_2$) is shorter than $S_2$ ($S_1$). This way, the shorter string will always be a prefix of the longer string; the longer string may have some extra characters which are the suffix of a string from the corresponding set. We will compute the following table: $L(i, j, p)$=the minimum length of the shorter string (among $S_1$ and $S_2$), such that: *(1)* if $p=1$, then $S_1$ is longer than $S_2$ - $S_1$ has the structure $S_2+q$ (+ is concatenation), where $q$ is the suffix of the string $i$ from the set $A$, starting at the position $j$ ($0 \leq j \leq len(i,1)$) ; we consider the positions of the strings to be indexed starting from $0$; *(2)* if $p=2$, then $S_2$ is longer than $S_1$ - $S_2$ has the structure $S_1+q$, where $q$ is the suffix of the string $i$ from the set $B$, which starts at the position $j$ ($0 \leq j \leq len(i,2)$)

We now have to solve a shortest path problem in a directed graph with $O(N \cdot (LMAX+1))$ vertices ($LMAX$=the maximum length of a string from $A$ or $B$), or $O$(sum of the lengths of all the strings from $A$ and $B$). In order to compute the neighbors of a vertex $(i,j,p)$, we will consider all the strings $i_1$ from the set $A$ (if $p=2$) or $B$ (if $p=1$). For every string $i_1$, we will verify if this string matches the string $i$ from the set $A$ (if $p=1$) or $B$ (if $p=2$), starting from the position $j$. Let's assume that the two strings match. Then, if $len(i_1,3-p) \leq len(i,p)-j$, then the vertex $(i,j,p)$ will have a directed edge towards the vertex $(i, j+len(i_1,3-p), p)$ - the cost of this edge will be $len(i_1,3-p)$; otherwise, the vertex $(i, j, p)$ will have a directed edge towards the vertex $(i_1, len(i_1, 3 - p) - len(i,p) + j, 3-p)$ – the cost of the edge will be $(len(i,p)-j)$.

For every vertex $(i,j,p)$ and every considered string $i_1$ we can test in $O(LMAX)$ time if the string $i_1$ matches the corresponding suffix from the string $i$ (starting at the position $j$). However, we can preprocess all this information using KMP. For each string $i$ from $A$ [$p=1$] ($B$ [$p=2$]) we consider every string $i_1$ from $B$ ($A$); we can compute in $O(LMAX)$ time all the positions $j$ from the string $i$ at which the string $i_1$ may start such that it matches the suffix of the string $i$ starting at $j$ ($j=j'-len(i_1,3-p)+1$ if $i_1$ matches $i$ on the positions $j,…,j'$, or $j=len(i,p)-j''$, where $j''$ is the length of any prefix of $i_1$ which is an ancestor in $i_1$'s failure tree of the longest prefix of $i_1$ matching the end of $i$). Thus, we can test a match in $O(1)$ amortized time for every tuple $(i,j,p,i_1)$.

The graph has $O(N \cdot (LMAX+1))$ vertices and $O(N^2 \cdot (LMAX+1))$ (directed) edges. Using Dijkstra's algorithm, the time complexity is $O(N^2 \cdot (LMAX+1)^2)$, or $O(N^2 \cdot (LMAX+1) \cdot log(N \cdot (LMAX+1)))$ (if we use binary heaps), or $O(N^2 \cdot (LMAX+1)+N \cdot (LMAX+1) \cdot log(N \cdot (LMAX+1)))$ (if we use Fibonacci heaps). At first, we will have $L(*,0,*)=0$ (the vertices $(i,0,p)$ are the source vertices; we can always have multiple source vertices if we insert them all at the beginning in the queue/binary heap/Fibonacci heap) and $L(*,j>0,*)= +\infty$. At the end of the algorithm, the length of the shortest string $S$ that can be written as both a concatenation of strings from $A$ and from $B$ is $min\{L(i, len(i,p), p)\}$. The string $S$ can be computed by tracing back the way we computed the values $L(i,j,p)$ (starting from the vertex which minimizes the length of $S$). If we hadn't cared for finding the shortest string $S$ with the given properties (and any string $S$ would have been ok), then we could have simply traversed the constructed graph (using DFS or BFS), starting from the vertices $(i,0,p)$ (we introduce all these vertices in the beginning in the BFS queue; in the DFS traversal case, we will start a new traversal from each such vertex, taking care not to traverse the vertices which were marked as traversed at previous DFS traversals or at the current DFS traversal). If we visit a vertex $(i,len(i,p),p)$, then a string $S$ with the given properties exists (and can be found by tracing back the path towards the source vertex/vertices). An application of this problem is the following. We have a set of strings $A$ and we want to compute the shortest palindrome that can be obtained as a concatenation of some strings from the set (any string may be used zero, one, or multiple times). We will construct the set $B$ as being formed of the strings from $A$, but reversed. Then, we will compute the same values as before. A vertex $(i,j,p)$ denotes a solution if the substring starting at the position $j$ from the string $i$ of the set corresponding to $p$ ($A$ for $p=1$, and $B$ for $p=2$) and ending at the end of the string is a palindrome (this substring may also have a length of $0$ or $1$). The answer will be $min\{2 \cdot L(i,j,p)+len(i,p)-j \mid$ the suffix of the string $i$ from the set $A$ ($B$) if $p=1$ ($p=2$), starting at the position $j$, is a palindrome$\}$. The concatenation of the strings from the set $B$ is reversed and attached at the end of the concatenation of the strings from $A$.

### B. Minimum Lexicographic Concatenation

We consider $N$ strings: $S(1), …, S(N)$. We want to sort these strings in some order $p(1), …, p(N)$, such that the string $Q=S(p(1))+…+S(p(N))$ (+ denotes the concatenation of two strings) is lexicographically minimum. Let $len(X)$ be the length of the string $X$. We will use any sorting algorithm for sorting "increasingly" the $N$ strings. When we need to compare two strings $S(i)$ and $S(j)$, we can decide that:
- if $S(i)+S(j) <_{lex} S(j)+S(i)$ then $S(i)$ "<" $S(j)$
- if $S(i)+S(j) >_{lex} S(j)+S(i)$ then $S(i)$ ">" $S(j)$
- if $S(i)+S(j)=S(j)+S(i)$ then: if $len(S(i)) \leq len(S(j))$ then $S(i)$ "≤" $S(j)$ else $S(i)$ ">" $S(j)$

IV. OPTIMAL LENGTH COMMON SUBSEQUENCES

A. *Longest Common Contiguous Subsequence*

We consider $N$ strings: $S(1), ..., S(N)$. We want to compute the longest contiguous substring which occurs at least $a(i)$ times ($0 \leq a(i)$; $1 \leq i \leq N$) in at least $F$ ($0 \leq F \leq N$) strings among the $N$ given (for the other strings $S(i)$, the substring may occur fewer than $a(i)$ times). We will construct a string $Z=S(1)\ \$_1\ S(2)\ \$_2\ ...\ \$_{N-1}\ S(N)$, structured as follows: the string $S(1)$, followed by the character $\$_1$, then followed by $S(2)$, then followed by $\$_2$, ..., then followed by $S(N)$. The characters $\$_1, ..., \$_{N-1}$ are distinct and they do not occur in any of the strings $S(i)$ ($1 \leq i \leq N$). We will construct a suffix array associated to $Z$, obtained by lexicographically sorting the suffixes of $Z$: $su(1), ..., su(|Z|)$ ($su(i)$ denotes the position of the first character from $Z$ of the corresponding suffix; $|Z|=len(Z)$). For each suffix $su(i)$ we know exactly to which string $S(i)$ its first character belongs, because of its position in $Z$ (the first $|S(1)|$ positions are marked as belonging to $S(1)$, the next position contains $\$_1$, the next $|S(2)|$ positions are marked as belonging to $S(2)$, and so on); the suffixes for which the first character is one of the characters $\$_1, ..., \$_{N-1}$ will present no interest. We will also compute the values $LCP(i)$=the length of the longest common prefix of the suffixes $su(i)$ and $su(i+1)$ ($1 \leq i \leq |Z|-1$). The suffix array and the array $LCP$ can be constructed in $O(|Z| \cdot log(|Z|))$ time. We will traverse the suffix array $su(*)$ with two pointers, *left* and *right*. Initially, we will have *left*=1 and *right*=0. We will maintain an array $x$, where $x(i)$=the number of occurrences of a suffix whose starting position belongs to $S(i)$, among the suffixes $su(j)$ with *left*$\leq j \leq$*right*; initially, $x(i)=0$ ($1 \leq i \leq N$). We will also maintain a counter *nok*=the number of strings $S(i)$ for which $a(i) \leq x(i)$. Initially, *nok* is equal to the number of indices $j$ for which $a(j)=0$ ($1 \leq j \leq N$). If *nok*$\geq F$ from the start, then there are at least $F$ values $a(i)=0$ and the answer is given by the string $S(q)$ having the maximum length. Another particular case occurs when there are exactly $F-1$ values $a(i)$ equal to $0$, and the other values are at least $1$ – in this case, the maximum length subsequence is the string $S(k)$ with the maximum length such that $a(k) \geq 1$. After removing the special cases, we will perform $|Z|$ steps. At the beginning of every step we will set *right*=*right*+1. If the suffix $su(right)$ belongs to a string $S(j)$, then we will set $x(j)=x(j)+1$; if $x(j)$ becomes equal to $a(j)$, then we set *nok*=*nok*+1. Then, we will increment the variable *left* by $1$ as long as one of the following conditions is true: *(1)* the first position of the suffix $su(left)$ belongs to no string $S(j)$ (i.e. it corresponds to a character $\$_q$); *(2)* the first position of $su(left)$ belongs to a string $S(j)$ for which $((x(j)>a(j))$ or $((x(j)=a(j))$ and $(nok>F)))$ : in this case we decrement $x(j)$ by $1$ first (and, if $x(j)$ becomes smaller than $a(j)$, we also decrement *nok* by $1$) and only after this will we increment *left* by $1$. After the (possible) changes of the variable *left*, we check if *nok*$\geq F$. If it is, then we will compute the value $W=min\{LCP(j)\ |\ left \leq j \leq right-1\}$ (if *left*=*right*, then $W$=the length of the suffix $su(left)$, i.e. $|Z|-$ $su(left)+1$; if the first character of $su(left)$ is a $\$_q$ character, then $W=-\infty$). We can compute these values in $O(1)$ time, using the *Range Minimum Query* (RMQ) technique, which requires a simple $O(|Z| \cdot log(|Z|))$ time preprocessing (or a more complicated $O(|Z|)$ time one). $W$ is the length of the longest substring of the strings $S(i)$ which satisfies the constraints, considering only the suffixes on the positions *left*, *left*+1, ..., *right*. We compare $W$ with the largest length $L$ found so far and we set $L=max\{W,L\}$ (initially, $L=0$). Finally, after all these operations, we can skip to the next step. The time complexity of the entire algorithm is $O(|Z| \cdot log(|Z|))$ (the preprocessing stage takes $O(|Z| \cdot log(|Z|))$ time; all the other $|Z|$ steps take $O(|Z|)$ time overall).

B. *Longest Common Non-Contiguous Subsequence*

We consider $K$ strings: $S(1), ..., S(K)$, composed of characters from an alphabet with $N$ symbols (numbered from $1$ to $N$); each position $j$ of a string $S(i)$ has a weight $wp(i,j) \geq 0$. A string $A$ is a (not necessarily contiguous) *subsequence* of another string $B$ if it can be obtained from $B$ by erasing $0$ or more characters from $B$. We want to compute a string $S$ which is a common subsequence of all the $K$ given strings and whose aggregate weight is maximum. The weight of a string is computed as an aggregate function $agg_1$ of the weights of its characters. The weight of a character on a position of the string is equal to an aggregate $agg_2$ of the weights of the positions in which the character matches each of the $K$ given strings; $agg_1$ is a non-decreasing function, defined for non-negative values (e.g., addition, max); $agg_2$ may be any function returning non-negative values (e.g. sum, multiplication, max, min). Each character $i$ occurs $num(j,i) \geq 1$ times in $S(j)$ (otherwise, we could remove the character from the alphabet and from every string which contains it). We know that the total number of tuples of positions $(p_1, ..., p_k)$, such that $S(1)(p_1)=S(2)(p_2)=...=S(K)(p_K)$ is at most *PMAX*. The positions of a string are numbered starting from $1$.

We will generate all the tuples $(p(1), ..., p(K))$ such that $S(1)(p(1))=...=S(K)(p(K))$. In order to do this, we will traverse every string $S(i)$ and, for every character $j$, we will construct a list $L(i,j)$ (initially, all these lists are empty). As we traverse the string $S(i)$ and we reach the character on a position $q$, we insert $q$ at the end of the list $L(i,S(i)(q))$. Thus, the elements of each list are added in increasing order. Then, we will generate all the tuples we mentioned, in lexicographic order. We will traverse, one at a time, the positions $q(1)$ from $S(1)$. For every position $q(1)$ we will consider, in order, every position $q(2)$ from $L(2, S(1)(q(1)))$; for every pair $(q(1), q(2))$ we consider, in order, every position $q(3)$ from $L(3, S(1)(q(1)))$, and so on, for every tuple $(q(1), ..., q(r))$ (with $r<K$) we traverse all the positions $q(r+1)$ from $L(r+1, S(1)(q(1)))$. Every time $q(K)$ takes a value, we generate a new tuple $(q(1), ..., q(K))$. The tuples are generated in lexicographic order. Every tuple $(q(1), ..., q(K))$ will have an associated weight $w(q(1), ..., q(K))=agg_2(wp(1,q(1)), ..., wp(K,q(K)))$. Then, we will traverse the tuples in the order we generated them and we will maintain a *(K-1)*-dimensional range tree. For every tuple *(p(1), ..., p(K))*, we will compute a weight *wmax(p(1), ..., p(K))=agg$_1$(w(p(1), ..., p(K)), max{wmax(p'(1), ..., p'(K)) | 1≤p'(i)<p(i) for every 1≤i≤K})*. In order to compute *wmax(p(1), ..., p(K))*, we will use the range tree for computing the maximum weight of a point *(p'(2), ..., p'(K))* in the range tree (with *p'(j)<p(j)* for every *2≤j≤K*); if no such point exists, the range tree will return the neutral value for the *agg$_1$* function (e.g. *0* for sum, *1* for multiplication, -∞ for max). After traversing all the tuples *(p(1), *, ..., *)* (i.e. with the same value of *p(1)*), we insert into the range tree a point *(p(2), ..., p(K))*, with the weight *wmax(p(1), ..., p(K))*, for each such tuple *(p(1), ..., p(K))*; these points will be considered when computing *wmax(*, ..., *)* for the next tuples. A point insertion and a (range) query can be performed in *O(log$^{K-1}$(PMAX))* time. Since all the points are known from the beginning, we can construct the range tree from the start for all the points, assigning to them weights equal to the neutral value of *agg$_1$*. Then, a *point insertion* will consist of changing the weight of a point in the range tree. However, the time complexities stay the same.

The final time complexity is *O(PMAX·log$^{K-1}$(PMAX))*. The maximum weight of a common (non contiguous) subsequence of the *K* strings is *max{wmax(p(1), ..., p(K))}*. Another interpretation of the problem is that we have a graph with *O(PMAX)* vertices; each vertex corresponds to a tuple *(p(1), ..., p(K))*. We have a directed edge from *(p'(1), ..., p'(K))* to *(p(1), ..., p(K))* if *p'(i)<p(i)* (for every *1≤i≤K*). Each vertex *(p(1), ..., p(K))* has a weight *w(p(1), ..., p(K))*. In this directed acyclic graph we need to find a path with maximum aggregate weight (using the *agg$_1$* function).

### C. Shortest Common Contiguous Non-Subsequence

Given a set of *R* strings, *St(1), ..., St(R)*, we want to find the shortest string *PS* which is not a contiguous subsequence of any of the *R* given strings. The strings' characters belong to an alphabet *A*. The characters of *PS* must belong to the same alphabet *A* (the characters of *A* are numbered from *1* to *|A|*). Obviously, such a string always exists, because we can always choose a string longer than any string *St(i)*. In [2], the authors simply mentioned that the problem can be solved in polynomial time and presented a trivial (but suboptimal solution) based on sorting all the *R* strings. If *|A|=1* then *len(PS)=1+max{len(St(i))|1≤i≤R}*. Otherwise, for every two consecutive strings in the sorted order, say *St(F)* and *St(G)*, we compute the shortest string *PS(F,G)* which is located between *St(F)* and *St(G)* in lexicographic order. In order to do this, we need to compute *LCP(F,G)*=the length of the longest common prefix of *St(F)* and *St(G)*. Then, *PS(F,G)* is constructed as follows: we take the prefix of length *LCP(F,G)* of *St(F)* and then: let $c_1$= *St(F)(LCP(F,G)+1)* and $c_2$=*St(G)(LCP(F,G)+1)*. We will consider *St(F)(y)=0*, if *y>len(St(F))* and *St(G)(y)=|A|+1*, if *y>len(St(G))*. If $c_1$+1≤$c_2$-1 then we add to *PS(F,G)* any character labeled with a number *lb* form the interval *[$c_1$+1,$c_2$-1]*. If, however, $c_1$+1=$c_2$, then we have two possibilities (and we will choose the one which leads to a shorter string). The first possibility occurs if $c_1$>0. We add to *PS(F,G)* the character $c_1$. Then, we need to compute the number of consecutive characters equal to *|A|* in *St(F)*, starting from position *LCP(F,G)+2*. Let *x=cnt(F, |A|, LCP(F,G)+2)* be this number. We add to *PS(F,G)* *x* characters *|A|* and then we add a character which is larger than *St(F)(LCP(F,G)+2+x)*. The second possibility occurs if $c_2$≤|A|. We add to *PS(F,G)* the character $c_2$. Then, we need to compute the number of consecutive characters equal to *1* in *St(G)*, starting from position *LCP(F,G)+2*. Let *x=cnt(G, 1, LCP(F,G)+2)* be this number. We add to *PS(F,G)* *x* characters *1* and then we add a character which is smaller than *St(G)(LCP(F,G)+2+x)*. We can tabulate the *cnt(*,*,*)* values as follows. For *y>len(St(U))* we have *cnt(U, c, y)=0*; for *1≤y≤len(St(U))* we have: if *(St(U)(y)=c)* then *cnt(U, c, y)=1+cnt(U, c, y+1)*; otherwise, *cnt(U, c, y)=0*. For completeness, we add an empty virtual string at the beginning of the sorted order and an empty string at the end of the sorted order. We can implement all these operations in *O(N·log(N))* time using the suffix arrays technique (*N*=the sum of the lengths of the *R* given strings, plus *R-1*): we construct a large string *S=ST(1) $_1$ St(2) $_2$ ... $_{R-1}$ St(R)*, where *$_1$, ..., $_{R-1}$* are different characters which are not part of *A*. Then, we can sort all the suffixes of *S* in *O(N·log(N))* time and compute the longest common prefix (LCP) of any two consecutive suffixes in the sorted order in *O(log(N))* time (by storing auxiliary information). Afterwards, we will maintain only the suffixes starting at the initial position of a string *ST(i)* and truncated, such that they do not contain *$_j$* characters. We can compute the LCP between any pair of suffixes in the original order by using the Range Minimum Query (RMQ) technique on the array of LCP values between consecutive suffixes in the sorted order; the length of the LCP between two truncated suffixes is reduced if it exceeds the length of any of them. Another method for computing the LCP of two suffixes starting at any positions *a* and *b* of *S* was mentioned to us by C. Negruşeri. We will compute a hash value *h(i)* for every position *i* of *S*: *h(0)=0* and *h(1≤i≤N)=hash(h(i-1), S(i), i)*. With these values, we will be able to efficiently compute a hash value for any (contiguous) substring *S(i:j)* (from the position *i* to *j*) of *S*. The hash values of two substrings with the same value will be identical. Then, we will binary search the LCP between *0* and *N-max{a,b}+1*. We have *LCP≥L* if *hashValue(a, a+L-1)=hashValue(b, b+L-1)* (with a high probability) and *LCP<L* otherwise. We can use *h(i)=($P_1^{i-1}$·S(i)+h(i-1)) mod $P_2$* and *hashValue(i,j)=((((h(j)-h(i-1)+$P_2$) mod $P_2$)·$P_1^{-(i-1)}$) mod $P_2$*, where $P_1$ and $P_2$ are two prime numbers and $P_1^{-1}$ is the modular multiplicative inverse of $P_1$ modulo $P_2$. After finding $P_1^{-1}$, we will compute and store the values $P_1^{-i}$=($P_1^{-(i-1)}$·$P_1^{-1}$) mod $P_2$ for *2≤i≤N*. We will present next a much simpler *O(N·log(N))* solution. One of the simplest solutions that comes to mind is to try every possible length *L* (starting from *1*) and, for each length *L*, generate all the *|A|$^L$* strings of length *L*. For each generated string, we verify if it is contained in the constructed string *S* (in *O(L+N)* time, using the KMP or Rabin-Karp methods). The time complexity of this solution is *O(N·|A|$^{len(PS)+1}$)*. Let's notice first that *PS*

cannot be too long. In order to have *len(PS)=L*, the string *S* must contain all the strings composed of *L-1* characters. The shortest string *S'* (with characters from the alphabet *A*) which contains all the strings composed of *Q* characters can be constructed using the following (well-known) algorithm. We construct a graph which contains a vertex for every string composed of *Q-1* characters. From every vertex *i* we have *|A|* outgoing (directed) edges. The $qe^{th}$ such edge is labeled with character *qe* ($1 \leq qe \leq |A|$). In order to find the vertex *j* into which this edge enters, we proceed as follows. Let *SQ(i)* be the *Q-1* character string associated to vertex *i*. We add character *qe* at the end of *SQ(i)* (obtaining the string *SQ'(i,qe)*) and then we keep only the last *Q-1* characters of *SQ'(i,qe)* (obtaining the string *SQ''(i,qe)*). Vertex *j* will be that vertex with *SQ(j)=SQ''(i,qe)*. In this graph, every vertex has the in-degree and the out-degree equal to *|A|*. We will compute an Euler cycle in this graph (as the graph is also connected). This cycle will contain all the $|A|^Q$ edges of the graph (there are $|A|^{Q-1}$ vertices, each of which has out-degree and in-degree *|A|*). In order to generate a string *S'* which contains all the strings composed of *Q* characters, we will choose an edge of the Euler cycle, which will be removed form the cycle. Then, we traverse the cycle starting from the next edge, until we reach the edge preceding the removed edge (at this moment, the entire cycle was traversed). For every traversed edge, we add its label at the end of *S'* (initially *S'* is empty). Let's assume that the removed edge was entering vertex *j*. Then we append at the beginning of *S'* the string *SQ(j)*. Thus, *S'* has $|A|^Q+Q-1$ characters. From the previously described algorithm we conclude that in order for the string *S* to contain all the strings composed of *L* characters, its length must be at least $|A|^L+L-1$. Thus, $L=O(log_{|A|}(N))$ and the initial algorithm has a time complexity of $O(N \cdot |A|^{log(N)+1})=O(N^2)$. However, the time complexity is too large. In order to reduce it, we will construct a trie (prefix tree), where we will add all the contiguous subsequences of the sequence *S*, composed of $L=log_{|A|}(N)+1$ characters. Building the trie takes $O(N \cdot log(N))$ time. Afterwards, we remove from the trie all the edges *(parent(q),q)* marked with $\$_j$ characters (and the subtrees rooted at these vertices *q*). Then, we will traverse all the nodes in the trie. Let's assume that we reached a node *q*, located on level *lev* ($0 \leq lev \leq L-1$). If node *q* does not have *|A|* sons, then we found a string composed of *lev+1* characters which is not included in *S*. This string is composed of the labels on the edges on the path from the root of the trie to node *q*, to which we add a character *c* from *A*, such that none of the edges connecting node *q* to one of its sons is labelled with *c*. *PS* will be the shortest string found this way. This stage of the algorithm takes *O(M)* time, where *M=O(N·log(N))* is the number of nodes in the trie.

V. CONSTRUCTING AND COUNTING STRINGS WITH SUBSTRING OCCURRENCE CONSTRAINTS

We consider an alphabet *A* with *M* characters, numbered from *1* to *M*. We are also given two lists $L_1$ and $L_2$ of strings composed of characters from the alphabet *A*. For each string *S(i)* from $L_2$ we are also given a set *Occ(i)* containing non-negative numbers not larger than *K*. We are interested in counting the total numbers of strings *SL* of a length *Len* ∈ *SLen* with characters from the alphabet *A* such that:
- none of the strings from $L_1$ occurs as a contiguous substring in *SL*
- the number *o(i)* of occurrences of each string *S(i)* from $L_2$ into *SL* is a number from *Occ(i)*; two occurrences of the same string *S(i)* may partially overlap.

We will start by constructing the deterministic finite automaton (DFA) of the strings from $L_1 \cup L_2$, like in [4]. Each state of the DFA which corresponds to the ending of a string from $L_1$ is marked as *forbidden*; the other states are not forbidden. For each non-forbidden state *q* we will compute the set *SE(q)* of the string indices *i* from $L_2$ such that the state *q* corresponds to matching the ending of the string *i*. We can do this like in [3], where they use the concept of *output links*. By following the output links starting at the state *q*, we will identify all the strings *i* whose ending matches the state *q* in *O(|SE(q)|)* time. We will now run a dynamic programming algorithm on the DFA. For each non-forbidden state *q* of the DFA and every tuple $(o(1), ..., o(|L_2|))$ (with $0 \leq o(i) \leq K$; $1 \leq i \leq |L_2|$) we will compute $Cnt(q, l, o(1), ..., o(|L_2|))$=the total number of strings *SL* of length *l* such that their suffix matches the string corresponding to the state *q* and each string *i* from $L_2$ occurs *o(i)* times in *SL* ($|L_2|$ denotes the number of strings in $L_2$; the strings from $L_2$ are numbered from *1* to $|L_2|$). In the initial state $q_0$ (which we assume it is not forbidden) of the DFA (corresponding to a void string) we have $Cnt(q_0, 0, o(1)=0, ..., o(i)=0, ..., o(|L_2|)=0)=1$ and for every other state $q \neq q_0$ we have $Cnt(q, 0, o(1)=*, ..., o(|L_2|)=*)=0$. We will compute these values in increasing order of *l* ($1 \leq l \leq max\{Len|Len \in SLen\}$). For each non-forbidden state *q* we will initialize $Cnt(q, l, o(1)=*, ..., o(i)=*, ..., o(|L_2|)=*)=0$. Then, we will consider all the directed edges entering *q* from other non-forbidden states *q'*. For each such state *q'* we will consider all the possible tuples $(o(1), ..., o(|L_2|))$ and we will increment $Cnt(q, l, o(1)+x(q,1), ..., o(i)+x(q,i), ..., o(|L_2|)+x(q,|L_2|))$ by the value $Cnt(q', l-1, o(1), ..., o(|L_2|))$. We have *x(q,i)=1* if $i \in SE(q)$ and *0*, otherwise ($1 \leq i \leq |L_2|$). The final result is the sum of the values $Cnt(q, Len, o(1), ... o(|L_2|))$ of the non-forbidden states *q*, where *Len* ∈ *SLen* and *o(i)* ∈ *Occ(i)* ($1 \leq i \leq |L_2|$). The time complexity is $O(Lmax \cdot K^{|L_2|} \cdot (V+E))$, where *Lmax=max{Len|Len* ∈ *SLen}*, *V* is the number of states of the DFA and *E* is the number of edges of the DFA.

A slightly more general problem is the following. We are given a DFA (with non-forbidden states only), in which we have an initial state and a set of final states (the initial state may also be final). We want to compute the number of strings of a length *Len* ∈ *SLen* that are accepted by the DFA (i.e. starting from the initial state and following the transitions corresponding to the characters of the string, we reach a final state). Moreover, some edges *q->q'* do not make us move to the next character of the string (i.e. they are *non-absorbing*). Every state *q* stores *4* lists of adjacent edges, for the following cases: *incoming/outgoing* and *absorbing/non-absorbing*. Moreover, within each list, it stores *M* sub-lists, one for each character of the alphabet (sub-list *c* contains the edges marked with *c* from the corresponding list). At first, we will process the DFA. For every character *c* from the alphabet, we consider *DFA'(c)*=the DFA containing all the states but only the non-absorbing edges marked with *c*. From each state *q* there is at most one outgoing edge marked with *c*. We will identify the cycles in *DFA'(c)* and remove from DFA and *DFA'(c)* all the edges which are part of a cycle in *DFA'(c)*. In order to identify the cycles, we consider that all the states are unmarked. Then, we consider every state *q* of *DFA'(c)*. If *q* is unmarked, we move along the edges, starting from the one going out of *q* (if any) and mark all the states we visit. If we get back to *q*, then we found a cycle. After removing these edges, *DFA'(c)* is a directed acyclic graph. We will compute a topological sort of *DFA'(c): ts(c,1), …, ts(c,V)*, such that all the states *q'* for which there is an edge *q'->q* (directed from *q'* to *q*) are located before *q* in the topological sort. Then, we will compute *Cnt(q,l)*=the number of strings of length *l* which can make the DFA reach the state *q*. We have *Cnt($q_0$,0)=1* and *Cnt(q≠$q_0$,0)=0* ($q_0$=the initial state). Then, for *l=1,…,Lmax* we perform the following steps: *(1)* for every character *c* we consider the states *ts(c,1), …, ts(c,V)* (in this order): we compute *Cnt'(ts(c,i),l,c)* as *Cnt(ts(c,i),l-1)* plus the sum of the values *Cnt'(ts(c,j),l,c)* (such that there is a non-absorbing edge marked with *c* from *ts(c,j)* to *ts(c,i)*); *(2)* we consider every state *q* and set *Cnt(q,l)* to the sum of the values *Cnt'(q',l,c)* (where *q'* is a state such that there is an absorbing directed edge from *q'* to *q* marked with the character *c*). The final answer is the sum of the values *Cnt(q,Len)*, where *q* is a final state and *Len*∈ *SLen*.

We will now consider a problem which is very similar to the first problem from this section. We are given the same input data, except that every string *i* from $L_2$ also has a weight *w(i)*. We want to construct a string *SL* with maximum weight, which obeys the same constraints from the previous problem. The weight of a string *SL* is the aggregate *agg* of the weights of each occurrence of a string *S(i)* from $L_2$ into *SL*. Additionally, for some of the strings *S(j)* from $L_2$ we do not care how many times they occur in *SL*. We will proceed as follows. At first, we compute the DFA of all the strings from $L_1 \cup L_2$ and we mark the forbidden states (as before). Then, for each non-forbidden state *q*, we compute *ws(q)*=the aggregate *agg* of the weights *w(i)* of the strings *i* from $L_2$ which are part of *SE(q)*. After this, we remove from $L_2$ and from all the sets *SE(\*)* those strings *i* for which we don't care how many times they occur in the optimal string *SL*. After removing these strings, we will renumber the remaining strings from *1* to *|$L_2$|* (we modify accordingly the numbering in $L_2$ and in every set *SE(\*)*). Then, we will run a dynamic programming algorithm which is very similar to the one from the first problem. We will compute *Wmax(q, l, o(1), …, o(|$L_2$|))*=the maximum weight of a string *SL* of length *l* whose suffix matches the string associated to the state *q* of the DFA and in which every (remaining) string *i* from $L_2$ occurs *o(i)* times and in which none of the strings from $L_1$ occur. We have *Wmax($q_0$,0,o(1)=…=o(|$L_2$|)=0)=0* and -∞ for all the other tuples with *l=0*. For each length *l* from *1* to *Lmax* (in increasing order) we consider all the non-forbidden states *q*. We initialize *Wmax(q, l, o(1)=\*, …, o(|$L_2$|)=\*)=-∞* and *Prev(q, l, o(1)=\*, …, o(|$L_2$|)=\*)=undefined*. Then, we consider all the directed edges entering *q* from other non-forbidden states *q'*. For each such state *q'* we consider all the tuples *(q', l-1, o(1), …, o(|$L_2$|))* such that *Wmax(q', l-1, o(1), …, o(|$L_2$|))>-∞* and if *agg(Wmax(q', l-1, o(1), …, o(|$L_2$|)), ws(q))>Wmax(q, l, o(1)+x(q,1), …, o(i)+x(q,i), …, o(|$L_2$|)+x(q,|$L_2$|))* then we set *Wmax(q, l, o(1)+x(q,1), …, o(i)+x(q,i), …, o(|$L_2$|)+x(q,|$L_2$|))=agg(Wmax(q', l-1, o(1), …, o(|$L_2$|)), ws(q))* and *Prev(q, l, o(1)+x(q,1), …, o(i)+x(q,i), …, o(|$L_2$|)+x(q,|$L_2$|))=q'*. In the end, we will compute *wm=max{Wmax(q, Len, o(1), …, o(|$L_2$|))| Len*∈ *SLen* and *o(i)*∈ *Occ(i) (1≤i≤|$L_2$|)}*. If *wm>-∞* then let *(q, Len, o(1), …, o(|$L_2$|))* be a tuple such that *Wmax(q, Len, o(1), …, o(|$L_2$|))=wm*. We will construct the string from the end towards the front. While *Len>0* do: *(1)* let *q'=Prev(q, Len, o(1), …, o(|$L_2$|))* ; *(2)* the character on the position *Len* of *SL* is the character on the directed edge from *q'* to *q* ; *(3)* *o(i)=o(i)-x(q,i) (1≤i≤|$L_2$|)* ; *(4) q=q'* ; *(5) Len=Len-1*.

## VI. RELATED WORK

The failure tree produced by the KMP algorithm was mentioned in [1], but was not used directly as it is. In [2], the authors studied several non-substring and non-subsequence problems, focusing on whether they belong to the P or NP class. They also gave a trivial, but inefficient polynomial algorithm for the shortest non-contiguous subsequence problem, while we presented an efficient solution. String concatenations are related to the shortest common supersequence problem [6]. A survey of longest common subsequence algorithms (of only two strings) was presented in [5]. The algorithms and data structures we used in this paper (e.g. KMP, trie, deterministic finite automaton) are well presented in several text books, like [3] and [4].

## VII. CONCLUSIONS AND FUTURE WORK

In this paper we presented new and practical algorithmic solutions for several string processing problems which are important in the automatic knowledge extraction and data mining fields. Our solutions make use of several standard, but efficient, algorithmic techniques and data structures. The presented algorithms can easily be implemented in many data mining and knowledge discovery applications.